\documentclass[twocolumn,showpacs,nopreprintnumbers,amsmath,pra,superscriptaddress]{revtex4}
\usepackage{amsmath,amssymb}

\usepackage{graphicx}

\begin{document}

\title{Vector Cooper Pairs and Coherent-Population-Trapping-like States in Ensemble of Interacting Fermions}
\author{A.V. Taichenachev}
\affiliation{Institute of Laser Physics SB RAS, Novosibirsk
630090, Russia\\e-mail: llf@laser.nsc.ru} \affiliation{Novosibirsk
State University, Novosibirsk 630090, Russia}

\author{V.I. Yudin}
\affiliation{Institute of Laser Physics SB RAS, Novosibirsk
630090, Russia\\e-mail: llf@laser.nsc.ru} \affiliation{Novosibirsk
State University, Novosibirsk 630090, Russia}

\begin{abstract}
Using the standard Hamiltonian of the BCS theory, we show that in
an ensemble of interacting fermions with the spin 1/2 there exist
coherent states $|NC\rangle$, which nullify the Hamiltonian of the
interparticle interaction (scattering). These states have an
analogy with the well-known in quantum optics  the coherent
population trapping (CPT) effect. The structure of these CPT-like
states corresponds to Cooper pairs with the total spin $S$=1. The
found states have a huge degree of degeneracy and carry a
macroscopic magnetic moment, that allows us to construct a new
model of the magnetism connected with the delocalized electrons in
metals (conductors). A principal possibility to apply the obtained
results to the superfluid $^3$He is also demonstrated.
\end{abstract}


\maketitle

\section{Introduction}

The effect of coherent population trapping (CPT) (see \cite{Alz,Ar1,Gr,Agap,Ar2} and references therein) is one of
nonlinear interference effects. Owing to number of its manifestations in different optical phenomena and its
practical applications CPT occupies one of leading place in modern laser physics, nonlinear and quantum optics.
For example, CPT is used in high-resolution spectroscopy \cite{Hemmer,Akulshin,Scully_1,Kitching}, nonlinear
optics of resonance media \cite{Harris,Kochar,Scully_2}, laser cooling \cite{Aspect1,Aspect2}, atom optics and
interferometry \cite{Marte,Weitz,Featonby}, physics of quantum information
\cite{mazets96,Fleschauer1,Fleschauer2,Liu,Phillips,Zibrov,Wal}.

In the case of classical resonant field the CPT theory has been
developed for a three-state model \cite{Ar2,Gr} as well as for
multi-level systems with account for the level degeneracy
\cite{Hioe,sm,Tumaikin,Taichenachev1,Taichenachev2}. Recently we
generalize this theory to the case of an ensemble of atoms
interacting with a quantized light field \cite{Taich,Taich2}. Note
also the presence of BCS-type states under the CPT conditions
\cite{Bolkart}.

From the very general point of view the essence of CPT can be
formulated as follows. Consider two quantum systems (particles or
fields) $A$ and $B$. The interaction between them is described by
the Hamiltonian  $\widehat{V}_{A-B}$. Then the CPT effect occurs
when there exists a non-trivial state $|NC\rangle$, which
nullifies the interaction:
\begin{equation}\label{CPT}
\widehat{V}_{A-B}|NC\rangle=0\,.
\end{equation}
In this state, obviously, the energy exchange between the systems
$A$ and $B$ is absent. However, information correlations of the
systems can be very strong, leading to important physical
consequences. Note that if the system $A$ is equivalent to the
system $B$, then the condition (\ref{CPT}) means the absence of
the field self-interaction  or of the interparticle interaction
\begin{equation}\label{CPT1}
\widehat{V}_{A-A}|NC\rangle=0\,.
\end{equation}
From this general viewpoint the standard CPT effect in the
resonant interaction of atoms with electromagnetic field is
deciphered as follows: $A$ and $B$ are ensembles of atoms and
resonant photons, respectively; $\widehat{V}_{A-B}=-(\hat{\bf
d}{\bf E})$ is the dipole interaction operator, and  $|NC\rangle$
is the dark state $|dark\rangle$:
\begin{equation}\label{dark}
-(\hat{\bf d}{\bf E})|dark\rangle=0\,,
\end{equation}
In the course of the interaction atoms are accumulated in the dark
state, after that they do not scatter light, and they are not
scattered by light. The information on various parameters of the
resonant field has been encoded in the state $|dark\rangle$
\cite{sm,Taich,Taich2}.

Our standpoint consists in the following. The CPT principle,
expressed by (\ref{CPT}) or (\ref{CPT1}), is sufficiently
universal and it can manifest in various branches of physics. The
significant progress in laser physics, spectroscopy, quantum and
nonlinear optics caused by the invention of the CPT effect
earnestly argues that such states should not be considered a
priori as a mathematical artefact, despite their uncommonness and
superficial paradoxicality. Thus, attempts to introduce CPT-like
states (when they are present, of course) into a description of
various phenomena from different branches of physics do not
contradict to the general physical principles and they are well
founded.

For the first time such a generalized approach to CPT has been developed in our early paper \cite{Tum}, where it
is pointed out that from a phenomenological viewpoint the CPT effect has a some likeness to the superconductivity.
In \cite{Tum} the following comparison is carried out: atoms and electromagnetic field from one side, electrons
and phonons form the other side. Indeed, a gas of atoms being in the dark state $|dark\rangle$ do not interact
with photons (see eq.(\ref{dark})), similarly to electrons in a superconducting state in solids, which are not
scattered by the phonon oscillations of a lattice. In the paper \cite{Tum} a hypothesis on the possibility of an
alternative (to the standard BCS theory \cite{Bar}) mechanism of superconductivity has been proposed. Namely, a
quantum system of electrons and phonons coupled by the interaction Hamiltonian $\widehat{V}_{e-phonon}$ was
considered. According to \cite{Tum}, the new mechanism of superconductivity could be based on the existence of
such a state $|NC\rangle$, which nullifies the interaction operator $\widehat{V}_{e-phonon}$:
\begin{equation}\label{NC1}
\widehat{V}_{e-phonon}|NC\rangle=0\,,
\end{equation}
analogously to eq.(\ref{dark}). However, an explicit form of the
state  $|NC\rangle$ was not found in \cite{Tum}.

In the present paper for the standard Hamiltonian of interparticle
interaction in the BCS model \cite{Bar} we find in explicit and
analytical form  CPT-like states of the type (\ref{CPT1}). In
contrast to the scalar Cooper pairs ($S$=$\,0$) in the standard
BCS theory such CPT-like states are formed by pairs with the spin
$S$=1, i.e. here a vector $p$-pairing takes place. However, these
states have a huge degree of degeneracy and due to this reason
they can not be used as a basis for new mechanism of the
superconductivity as it has been suggested in our previous papers
\cite{Tum, TY}. Nevertheless, the existence of such CPT-like
states can lead to serious consequences, because these states
carry a macroscopic magnetic moment, that allows a principal
possibility to describe a new approach to the magnetism connected
with the delocalized electrons in metals (conductors). Apart from
these, the obtained results can be related to the description of
the superfluid phase of $^3$He, which, as is known, appears due to
the formation of Cooper pairs with the spin $S$=1.

\section{Ensemble of fermions in a finite volume}
Consider an ensemble of ${\cal N}$ fermions confined to a volume  $V$, where ${\cal N}$=$n$$V$, and $n$ is the
density of particles. We will use the standard BCS Hamiltonian \cite{Bar}:
\begin{equation}\label{H}
\widehat{H}_{BCS}=\widehat{H}_0+\widehat{W}\,.
\end{equation}
The Hamiltonian of free particles can be written as:
\begin{equation}\label{H_0}
\widehat{H}_0=\sum_{s,{\bf k}}\varepsilon^{}_{\bf
k}\,\hat{a}^{\dag}_{s{\bf k}}\hat{a}^{}_{s{\bf k}}\,,
\end{equation}
where $\hat{a}^{\dag}_{s {\bf k}}$($\hat{a}^{}_{s {\bf k}}$) is
the creation (annihilation) operator of Fermi particle in the
state with wavevector ${\bf k}$ and spin projection
$s=\uparrow,\downarrow$, and $\varepsilon^{}_{k}$ is the energy of
this state. These operators satisfy the following anticommutator
relationships:
\begin{eqnarray}\label{Ferm comm}
&& \hat{a}^{\dag}_{s {\bf k}}\hat{a}^{}_{s' {\bf
k'}}+\hat{a}^{}_{s' {\bf k'}}\hat{a}^{\dag}_{s {\bf
k}}=\delta^{}_{ss'}\delta^{}_{\bf kk'}, \nonumber\\
&& \hat{a}^{}_{s {\bf k}}\hat{a}^{}_{s' {\bf k'}}+\hat{a}^{}_{s'
{\bf k'}}\hat{a}^{}_{s {\bf
k}}=0, \nonumber\\
&& \hat{a}^{\dag}_{s {\bf k}}\hat{a}^{\dag}_{s' {\bf
k'}}+\hat{a}^{\dag}_{s' {\bf k'}}\hat{a}^{\dag}_{s {\bf k}}=0\,.
\end{eqnarray}
The interaction between particles is described by the Hamiltonian
coupling particles with opposite momenta and spin:
\begin{eqnarray}\label{W}
&&\widehat{W}=\frac{g}{V}\sum_{{\bf k}_{1,2}\in {\cal D}_F}G({\bf k}_1,{\bf k}_2)\,\hat{a}^{\dag}_{\uparrow {\bf
k}_1}\hat{a}^{\dag}_{\downarrow -{\bf k}_1} \hat{a}^{}_{\downarrow -{\bf k}_2}\hat{a}^{}_{\uparrow {\bf k}_2}\,,\\
&&{\cal D}_F :\;\; k_F-\Delta\le |{\bf k}|\le
k_F+\Delta\,.\nonumber
\end{eqnarray}
Only particles with wavevectors in the thin layer of the width $2\Delta$ around the Fermi surface (see in
Fig.1a), having the radius $k_F$ ($\Delta$$\ll$$\,k_F$), are involved in the interaction. This subset in the
wavevector space will be referred to as ${\cal D}_F$. If even one of the vectors ${\bf k}_{1,2}$ does not belong
to the subset ${\cal D}_F$, then $G$(${\bf k}_1,$${\bf k}_2$)=0. The sign of the interaction constant $g$ in
(\ref{W}) governs the attraction ($g$$<$0) or repulsion ($g$$>$0) between particles. The formfactor $G$(${\bf
k}_1,$${\bf k}_2$) obeys to the general symmetry condition
\begin{equation}\label{G}
G({\bf k}_1,{\bf k}_2)=G(-{\bf k}_1,{\bf k}_2)=G({\bf k}_1,-{\bf
k}_2)\,.
\end{equation}
In this case the Hamiltonian $\widehat{W}$ consists of the
quadratic operator constructions:
\begin{eqnarray}\label{aa}
&&\hat{a}^{\dag}_{\uparrow {\bf k}}\hat{a}^{\dag}_{\downarrow
-{\bf k}}+\hat{a}^{\dag}_{\uparrow -{\bf
k}}\hat{a}^{\dag}_{\downarrow {\bf k}}=\hat{a}^{\dag}_{\uparrow
{\bf k}}\hat{a}^{\dag}_{\downarrow -{\bf
k}}-\hat{a}^{\dag}_{\downarrow {\bf k}}\hat{a}^{\dag}_{\uparrow -{\bf k}}\,,\nonumber\\
&&\hat{a}^{}_{\uparrow {\bf k}}\hat{a}^{}_{\downarrow -{\bf
k}}+\hat{a}^{}_{\uparrow -{\bf k}}\hat{a}^{}_{\downarrow {\bf
k}}=\hat{a}^{}_{\uparrow {\bf k}}\hat{a}^{}_{\downarrow -{\bf
k}}-\hat{a}^{}_{\downarrow {\bf k}}\hat{a}^{}_{\uparrow -{\bf
k}}\,,
\end{eqnarray}
which are antisymmetrical with respect to the spin variables
($\uparrow$,$\downarrow$) and, consequently, they are scalar
constructions. Thus, the operator $\widehat{W}$ has the invariant
form (\ref{W}) independently of the direction of the quantization
axis $z$, with respect to which the spin projections
($\uparrow$,$\downarrow$) are determined. Thus, the relationship
(\ref{G}) is the consequence of the invariance of the Hamiltonian
(\ref{W}) with respect to the choice of the quantization axis $z$.
It is usually assumed that $G$(${\bf k}_1,$${\bf k}_2$)=1 at ${\bf
k}_{1,2}$$\in$${\cal D}_F$.

Recall, that, according to the standard conceptions of the
description of the conductivity electrons in metals, the model
Hamiltonian (\ref{W}) is governed by the interaction of electrons
with phonons of lattice and Coulomb repulsion between electrons.

It should be especially noted that since the interaction
Hamiltonian (\ref{W}) consists of the scalar (with respect to the
spin) constructions (\ref{aa}), then it was widely recognized that
the standard mathematical BCS model describes only the scalar
pairing of fermions. However, as it will be shown below, in the
general case it is not so, i.e. the Hamiltonian (\ref{W}) can
describe the vector pairing as well.

\begin{figure}[t]
\centerline{\scalebox{0.4}{\includegraphics{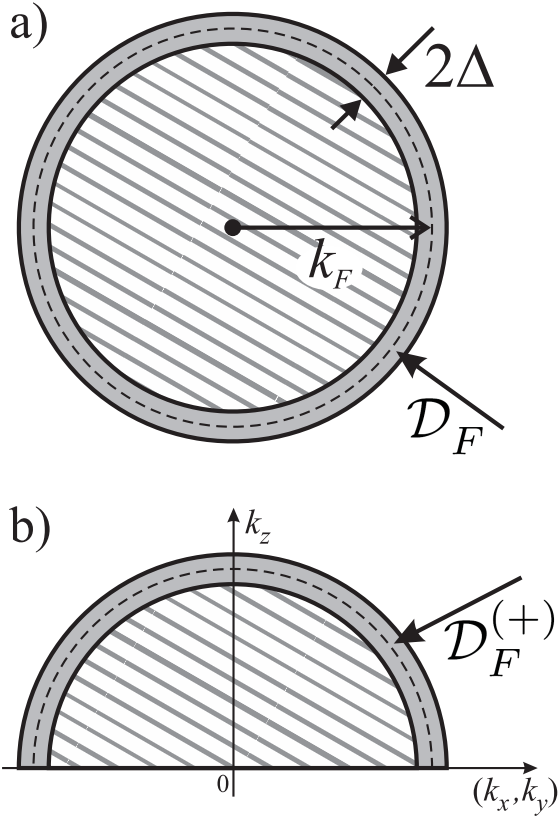}}}\caption{Illustrations: a) thin spherical layer ${\cal
D}_F$ with the width 2$\Delta$ around the Fermi surface with the radius $k_F$; b) upper hemispherical layer
${\cal D}_F^{(+)}$ with $k_z$$>$0.}
\end{figure}

\section{CPT-like states}
It turns out that the operator (\ref{W}) allows the existence of the CPT-like states $|NC\rangle$, obeying the
condition
\begin{equation}\label{W_NC1}
\widehat{W}|NC\rangle=0\,.
\end{equation}
Let us build up these states. Consider first the following
operator construction:
\begin{equation}\label{constr}
\hat{b}^{\dag}_{\bf k}(\xi)=(\hat{a}^{\dag}_{\uparrow {\bf
k}}\hat{a}^{\dag}_{\downarrow -{\bf k}}+\xi\,
\hat{a}^{\dag}_{\uparrow -{\bf k}}\hat{a}^{\dag}_{\downarrow {\bf
k}})/\sqrt{2}\,,
\end{equation}
which generates two-particle coupled states with opposite wavevectors ${\bf k}$ and $-{\bf k}$, and with the zero
projection of the total spin; the parameter $\xi$ is arbitrary number. Using the standard anticommutator rules
for fermionic operators $\hat{a}^{\dag}_{s {\bf k}}$ and $\hat{a}^{}_{s {\bf k}}$, and the property (\ref{G}), we
calculate the commutator:
\begin{eqnarray}\label{comm1}
&&\left[\widehat{W},\,\hat{b}^{\dag}_{\bf k}(\xi)\right]=\frac{g}{V\sqrt{2}}\sum_{{\bf k}_{1}\in {\cal
D}_F}G({\bf k}_1,{\bf k})\,\hat{a}^{\dag}_{\uparrow
{\bf k}_1}\hat{a}^{\dag}_{\downarrow -{\bf k}_1}\\
&&(1+\xi-\xi\hat{a}^{\dag}_{\uparrow -{\bf k}}\hat{a}^{}_{\uparrow
-{\bf k}}- \xi\hat{a}^{\dag}_{\downarrow {\bf
k}}\hat{a}^{}_{\downarrow {\bf k}}-\hat{a}^{\dag}_{\uparrow {\bf
k}}\hat{a}^{}_{\uparrow {\bf k}}- \hat{a}^{\dag}_{\downarrow -{\bf
k}}\hat{a}^{}_{\downarrow -{\bf k}})\nonumber .
\end{eqnarray}
As is seen, when $\xi=-1$ this commutator has the specific form, where all summands are finished by the
annihilation operators $\hat{a}^{}_{\uparrow \pm{\bf k}}$ and $\hat{a}^{}_{\downarrow \pm{\bf k}}$. Therefore we
define now the basic operator construction $\hat{\gamma}^{\dag}_{0,\bf k}$:
\begin{eqnarray}\label{base}
&&\hat{\gamma}^{\dag}_{0,\bf k}\equiv\hat{b}^{\dag}_{\bf
k}(-1)=(\hat{a}^{\dag}_{\uparrow {\bf
k}}\hat{a}^{\dag}_{\downarrow -{\bf k}}-\hat{a}^{\dag}_{\uparrow
-{\bf k}}\hat{a}^{\dag}_{\downarrow {\bf
k}})/\sqrt{2}=\nonumber \\
&&(\hat{a}^{\dag}_{\uparrow {\bf k}}\hat{a}^{\dag}_{\downarrow
-{\bf k}}+\hat{a}^{\dag}_{\downarrow {\bf
k}}\hat{a}^{\dag}_{\uparrow -{\bf k}})/\sqrt{2}\,,
\end{eqnarray}
which is symmetric on the spin variables $\uparrow$,$\downarrow$.
For this construction the commutator (\ref{comm1}) takes the form:
\begin{eqnarray}\label{comm2}
&&\left[\widehat{W},\,\hat{\gamma}^{\dag}_{0,\bf k}\right]=\frac{g}{V\sqrt{2}}\sum_{{\bf k}_{1}\in {\cal
D}_F}G({\bf k}_1,{\bf k})\,\hat{a}^{\dag}_{\uparrow
{\bf k}_1}\hat{a}^{\dag}_{\downarrow -{\bf k}_1}\nonumber\\
&&(\hat{a}^{\dag}_{\uparrow -{\bf k}}\hat{a}^{}_{\uparrow -{\bf
k}}+ \hat{a}^{\dag}_{\downarrow {\bf k}}\hat{a}^{}_{\downarrow
{\bf k}}-\hat{a}^{\dag}_{\uparrow {\bf k}}\hat{a}^{}_{\uparrow
{\bf k}}- \hat{a}^{\dag}_{\downarrow -{\bf
k}}\hat{a}^{}_{\downarrow -{\bf k}})\,.
\end{eqnarray}
Apart from the operator $\hat{\gamma}^{\dag}_{0,\bf k}$ there
exist the other two quadratic in the operators $\hat{a}^{\dag}_{s
\pm{\bf k}}$ constructions with the zero total momentum, for which
the commutator with the operator $\widehat{W}$ is finished from
the right side by the annihilation operators $\hat{a}^{}_{s
\pm{\bf k}}$. These construction are:
\begin{equation}\label{base1}
\hat{\gamma}^{\dag}_{+1,\bf k}=\hat{a}^{\dag}_{\uparrow {\bf k}}\hat{a}^{\dag}_{\uparrow -{\bf k}}\,,\quad
\hat{\gamma}^{\dag}_{-1,\bf k}=\hat{a}^{\dag}_{\downarrow {\bf k}}\hat{a}^{\dag}_{\downarrow -{\bf k}}\,,
\end{equation}
for them the following commutator relations are fulfilled:
\begin{eqnarray}\label{comm_p}
\left[\widehat{W},\,\hat{\gamma}^{\dag}_{+1,\bf k}\right]=&&\frac{g}{V}\sum_{{\bf k}_{1}\in {\cal D}_F}G({\bf
k}_1,{\bf k})\,\hat{a}^{\dag}_{\uparrow
{\bf k}_1}\hat{a}^{\dag}_{\downarrow -{\bf k}_1}\nonumber\\
&&(-\hat{a}^{\dag}_{\uparrow -{\bf k}}\hat{a}^{}_{\downarrow -{\bf k}}+ \hat{a}^{\dag}_{\uparrow {\bf
k}}\hat{a}^{}_{\downarrow {\bf k}})\,,
\end{eqnarray}
\begin{eqnarray}\label{comm_m}
\left[\widehat{W},\,\hat{\gamma}^{\dag}_{-1,\bf k}\right]=&&\frac{g}{V}\sum_{{\bf k}_{1}\in {\cal D}_F}G({\bf
k}_1,{\bf k})\,\hat{a}^{\dag}_{\uparrow
{\bf k}_1}\hat{a}^{\dag}_{\downarrow -{\bf k}_1}\nonumber\\
&&(\hat{a}^{\dag}_{\downarrow -{\bf k}}\hat{a}^{}_{\uparrow -{\bf k}}- \hat{a}^{\dag}_{\downarrow {\bf
k}}\hat{a}^{}_{\uparrow {\bf k}})\,.
\end{eqnarray}
The expressions (\ref{comm2}), (\ref{comm_p}), and (\ref{comm_m})
are crucial for the building up the CPT-like states (\ref{W_NC1}).

A set of the three operators  $\hat{\gamma}^{\dag}_{q,\bf k}$
($q$=0,$\pm 1$) for the given ${\bf k}$ constitutes an invariant
(with respect to the choice of the quantization axis $z$) subspace
and it describes the three orthogonal components of the particle
spin $S$=1 (i.e. it corresponds to the vector particle). These
components correspond to the spin projections (0,$\pm\hbar$) onto
the $z$ axis. Thus, here we deal with the vector pairing of
$p$-type, contrary to the scalar pairing of $s$-type ($S$=0).

Due to the obvious relationship
\begin{equation}\label{k-k}
\hat{\gamma}^{\dag}_{q,-\bf k}=-\hat{\gamma}^{\dag}_{q,\bf
k}\,,\quad (q=0,\pm 1)
\end{equation}
the operators $\hat{\gamma}^{\dag}_{q,\bf k}$, defined on the
spherical layer ${\bf k}$$\in$${\cal D}_F$, are not independent.
Therefore instead of the ${\cal D}_F$ we define a hemispherical
layer in the wavevector space. For example, choose the upper
hemispherical layer ${\cal D}_F^{(+)}$ (see Fig.1b), consisting of
vectors ${\bf k}$$\in$${\cal D}_F$ with positive projections on
the axis $z$ ($k_z$$>$$\,0$) only. Now the operators
$\hat{\gamma}^{\dag}_{q,\bf k}$, defined for vectors ${\bf
k}$$\in$${\cal D}_F^{(+)}$ are independent. Note, that the
introduction of the hemispherical layer in the wavevector space
plays an auxiliary role, reducing some notations. The concrete
choice of the hemispherical layer do not effect on the following
results.

Let us demonstrate the method of the construction of the CPT-like
states (\ref{W_NC1}) using a concrete example. Consider the
operator construction of the following form
\begin{equation}\label{Psi0}
\widehat{\Psi}_{NC}(...\hat{\gamma}^{\dag}_{0,\bf
k}...)=\prod_{{\bf k}\in {\cal
D}_F^{(+)}}\hat{\gamma}^{\dag}_{0,\bf k}\,,
\end{equation}
consisting only of the operators $\hat{\gamma}^{\dag}_{0,\bf k}$, which correspond to the zero spin projections
of the vector particles (\ref{base}). This construction  acts on the upper hemispherical layer ${\cal D}_F^{(+)}$
(for each ${\bf k}$ the operator $\hat{\gamma}^{\dag}_{0,\bf k}$ is used, at most, once). Obviously, the order of
multipliers can be arbitrary, because [$\hat{\gamma}^{\dag}_{q,\bf k}$$,\,$$\hat{\gamma}^{\dag}_{q',\bf k'}$]=0.
Let us factor out arbitrary operator $\hat{\gamma}^{\dag}_{0,\bf k'}$ in (\ref{Psi0}) from the product $\Pi$ and
then act by the operator $\widehat{W}$ on $\widehat{\Psi}_{NC}(...\hat{\gamma}^{\dag}_{0,\bf k}...)$:
\begin{eqnarray}\label{W_Psi}
&&\widehat{W}\widehat{\Psi}_{NC}(...\hat{\gamma}^{\dag}_{0,\bf
k}...)=\widehat{W}\hat{\gamma}^{\dag}_{0,\bf k'}\prod_{{\bf k}\neq
{\bf k'}
}\hat{\gamma}^{\dag}_{0,\bf k}=\nonumber \\
&&\left(\hat{\gamma}^{\dag}_{0,\bf
k'}\widehat{W}+[\widehat{W},\,\hat{\gamma}^{\dag}_{0,\bf
k'}]\right)\prod_{{\bf k}\neq {\bf k'} }\hat{\gamma}^{\dag}_{0,\bf
k}.
\end{eqnarray}
Since under the sign $\Pi$ in (\ref{W_Psi}) the creation operators with wavevectors $\pm{\bf k'}$ are absent,
then, as is follows from eq.(\ref{comm2}), the commutator $[\widehat{W},\,\hat{\gamma}^{\dag}_{0,\bf k'}]$ can be
moved to the right side through the product $\Pi$. As a result, the expression (\ref{W_Psi}) can be written as:
\begin{eqnarray}\label{W_Psi1}
&&\widehat{W}\widehat{\Psi}_{NC}(...\hat{\gamma}^{\dag}_{0,\bf k}...)=\nonumber \\
&&\hat{\gamma}^{\dag}_{0,\bf k'}\widehat{W}\prod_{{\bf k}\neq {\bf
k'} }\hat{\gamma}^{\dag}_{0,\bf k}+\left(\prod_{{\bf k}\neq {\bf
k'} }\hat{\gamma}^{\dag}_{\bf
k}\right)[\widehat{W},\,\hat{\gamma}^{\dag}_{0,\bf k'}]\,.
\end{eqnarray}
Let us consider also the operator construction
\begin{equation}\label{Ferm1}
\widehat{\Phi}(\Delta)=\prod_{|{\bf k}|<
(k^{}_F-\Delta)}\hat{a}^{\dag}_{\uparrow{\bf
k}}\hat{a}^{\dag}_{\downarrow{\bf k}}\,,
\end{equation}
which, acting on the vacuum $|0\rangle$, generates the state, corresponding to the completely occupied sphere
with the radius ($k_F-\Delta$) in the wavevector space (in Fig.1a it corresponds to the inner sphere shaded by
skew lines). The following commutator relationships are evident:
\begin{equation}\label{comm_F}
[\widehat{W},\, \widehat{\Phi}(\Delta)]=0;\quad \left[
[\widehat{W},\,\hat{\gamma}^{\dag}_{q,\bf k}],\,
\widehat{\Phi}(\Delta)\right]=0\quad ({\bf k}\in {\cal D}_F)\,,
\end{equation}
because in the operator  $\widehat{\Phi}(\Delta)$ (see (\ref{Ferm1})) only the wavevectors  $|{\bf
k}|$$<$$(k_F$$-\Delta)$ are used. These vectors do not belong to the upper layer ${\cal D}_F$ where the operators
$\widehat{W}$ and $\hat{\gamma}^{\dag}_{q,\bf k'}$ act.

Let us prove that the state $|NC,...\hat{\gamma}^{\dag}_{0,\bf k}...\rangle$, nullifying the interaction
(\ref{W_NC1}), has the form
\begin{equation}\label{NC_constr}
|NC,...\hat{\gamma}^{\dag}_{0,\bf
k}...\rangle=\widehat{\Psi}_{NC}(...\hat{\gamma}^{\dag}_{0,\bf
k}...)\widehat{\Phi}(\Delta)|0\rangle\,.
\end{equation}
Acting on this state by the operator $\widehat{W}$, and taking into account the relationships (\ref{W_Psi1}) and
(\ref{comm_F}), one can obtain:
\begin{eqnarray}\label{W_NCconstr}
&&\widehat{W}\widehat{\Psi}_{NC}(...\hat{\gamma}^{\dag}_{0,\bf
k}...)\widehat{\Phi}(\Delta)|0\rangle =\hat{\gamma}^{\dag}_{\bf
k'}\widehat{W}\left(\prod_{{\bf k}\neq {\bf k'}
}\hat{\gamma}^{\dag}_{0,\bf k}\right)\widehat{\Phi}(\Delta)|0\rangle\nonumber\\
&&+\left(\prod_{{\bf k}\neq {\bf k'} }\hat{\gamma}^{\dag}_{\bf
k}\right)\widehat{\Phi}(\Delta)\,[\widehat{W},\,\hat{\gamma}^{\dag}_{0,\bf
k'}]\,|0\rangle\,.
\end{eqnarray}
However, since the commutator (\ref{comm2}) is finished from the right side by the annihilation operators, then
[$\widehat{W}$$,\,$$\hat{\gamma}^{\dag}_{0,\bf k'}$]$|0\rangle=0$. Thus, from (\ref{W_NCconstr}) we have
\begin{eqnarray}\label{W_NCcomm}
&&\widehat{W}\widehat{\Psi}_{NC}(...\hat{\gamma}^{\dag}_{0,\bf
k}...)\widehat{\Phi}(\Delta)|0\rangle
=\nonumber\\
&&\hat{\gamma}^{\dag}_{0,\bf k'}\widehat{W}\left(\prod_{{\bf
k}\neq {\bf k'} }\hat{\gamma}^{\dag}_{0,\bf
k}\right)\widehat{\Phi}(\Delta)|0\rangle\,,
\end{eqnarray}
From this equation we see that it is possible to change the order of the sequence of $\widehat{W}$ and any
operator $\hat{\gamma}^{\dag}_{0,\bf k}$. Proceeding this consideration step by step and taking into account
(\ref{comm_F}), we obtain:
\begin{eqnarray}\label{W_NC_fin}
&&\widehat{W}|NC,...\hat{\gamma}^{\dag}_{0,\bf k}...\rangle \equiv
\widehat{W}\widehat{\Psi}_{NC}(...\hat{\gamma}^{\dag}_{0,\bf
k}...)\widehat{\Phi}(\Delta)|0\rangle =\nonumber\\
&&\widehat{\Psi}_{NC}(...\hat{\gamma}^{\dag}_{0,\bf k}...)
\widehat{W}\widehat{\Phi}(\Delta)|0\rangle=\nonumber\\
&&\widehat{\Psi}_{NC}(...\hat{\gamma}^{\dag}_{0,\bf
k}...)\widehat{\Phi}(\Delta)\widehat{W}|0\rangle=0\,.
\end{eqnarray}
Here the last transformation to zero is obvious, because the operator $\widehat{W}$ (see (\ref{W})) is finished
from the right side by the annihilation operators $\hat{a}^{}_{s\pm {\bf k}}$. Thus, we prove rigorously that the
state (\ref{NC_constr}) nullifies the interparticle interaction (scattering), i.e. it obeys the equation
(\ref{W_NC1}).

Consider now instead of the particular construction $\widehat{\Psi}_{NC}(...\hat{\gamma}^{\dag}_{0,\bf k}...)$
(see (\ref{Psi0})) the more general operator construction:
\begin{equation}\label{Psi_j}
\widehat{\Psi}_{NC}(...\hat{\gamma}^{\dag}_{q^{}_{\bf k},\bf k}...)=\prod_{{\bf k}\in {\cal
D}_F^{(+)}}\hat{\gamma}^{\dag}_{q^{}_{\bf k},\bf k}\,,\quad (q^{}_{\bf k}=0,\pm 1)\,,
\end{equation}
where the operators $\hat{\gamma}^{\dag}_{0,\bf k}$ and $\hat{\gamma}^{\dag}_{\pm 1,\bf k}$ can be used in
different ways. Each wave vector ${\bf k}$ should appear only one time (independent of the value $q^{}_{\bf k}$).
Performing mathematical calculations analogous to aforecited and taking into account the commutator relations
(\ref{comm2}), (\ref{comm_p}), (\ref{comm_m}), and (\ref{comm_F}), it can be easily seen, that any state of the
form
\begin{equation}\label{NC_j}
|NC,...\hat{\gamma}^{\dag}_{q^{}_{\bf k},\bf
k}...\rangle\equiv\widehat{\Psi}_{NC}(...\hat{\gamma}^{\dag}_{q^{}_{\bf k},\bf
k}...)\widehat{\Phi}(\Delta)|0\rangle
\end{equation}
obeys the equation (\ref{W_NC1}), i.e. it is a CPT-like state.

From the construction (\ref{Psi_j}) it follows that the states
$|NC,...\hat{\gamma}^{\dag}_{q^{}_{\bf k},\bf k}...\rangle$ have
the form of wave function of a system of non-interacting particles
with the spin 1 and they are characterized by a set of indices
$\{...,q^{}_{\bf k},...\}$. Let us denote the number of fermions
in the spherical layer ${\cal D}_F$ as $\overline{\delta {\cal
N}}$. Consequently, the number of vector pairs will be
$\overline{\delta {\cal N}}/2$. Then the number of different
CPT-like states $|NC,...\hat{\gamma}^{\dag}_{q^{}_{\bf k},\bf
k}...\rangle$ is equal to $3^{\overline{\delta {\cal N}}/2}$. By
its sense the number $\overline{\delta {\cal N}}$ equals to the
number of electrons in the spherical layer
($k^{}_F$$-\Delta$)$\leq$$k$$\leq$$k^{}_F$ at the dense packing in
the Fermi sphere in the absence of the interaction (\ref{W}). In
the case of $\Delta$$\ll$$k^{}_F$ we have the following
relationship:
\begin{equation}\label{dN}
\frac{\overline{\delta {\cal N}}}{{\cal N}}\approx 3\,\frac{\Delta}{k^{}_F}\,.
\end{equation}
It should be noted the presence of the construction $\widehat{\Phi}(\Delta)$ in (\ref{NC_j}) is necessary from the
physical point of view, since the form of the interaction Hamiltonian (\ref{W}), according to \cite{Bar}, is a
consequence of almost completely  occupied Fermi sphere. Thus, physically significant states should differ from
the ideal Fermi state $|F\rangle$:
\begin{equation}\label{Fermi}
|F\rangle=\left(\prod_{|{\bf k}|\le
k_F}\hat{a}^{\dag}_{\uparrow{\bf k}}\hat{a}^{\dag}_{\downarrow{\bf
k}}\right)|0\rangle
\end{equation}
only in a small region nearby the Fermi sphere. For the state (\ref{NC_j}) this difference is described by the
construction $\widehat{\Psi}_{NC}$ (\ref{Psi_j}), acting in the thin layer ${\cal D}_F$ around the Fermi surface
in the wavevector space.

As is easily seen, any state $|NC,...\hat{\gamma}^{\dag}_{q^{}_{\bf k},\bf k}...\rangle$ is an eigenstate for the
unperturbed Hamiltonian $\widehat{H}_0$ and, consequently, for the total Hamiltonian $\widehat{H}_{BCS}$:
\begin{eqnarray}\label{H_NC}
&&\widehat{H}_{BCS}\,|NC,...\hat{\gamma}^{\dag}_{q^{}_{\bf k},\bf k}...\rangle=\nonumber \\
&&\widehat{H}_0\, |NC,...\hat{\gamma}^{\dag}_{q^{}_{\bf k},\bf
k}...\rangle=E_{NC}\, |NC,...\hat{\gamma}^{\dag}_{q^{}_{\bf
k},\bf k}...\rangle \,.
\end{eqnarray}
In the case of quadratic dispersion law
\begin{equation}\label{k2}
\varepsilon^{}_{\bf k}=\frac{(\hbar {\bf k})^2}{2m}
\end{equation}
the eigenenergy is
\begin{equation}\label{E_NC}
E_{NC}=E_{F}+\Delta E_{NC}\,,
\end{equation}
where ${E}_F$ is the energy of an ideal Fermi-sphere:
\begin{equation}\label{E_F}
E_F=\frac{3(\hbar k^{}_F)^2{\cal N}}{10\, m}\,,
\end{equation}
and $\Delta E_{NC}$ is the relatively small ($\Delta E_{NC}$$\ll$$E_F$) positive contribution to the energy
\begin{equation}\label{DE_NC}
\Delta E_{NC}=\frac{3(\hbar k^{}_F)^2{\cal N}}{m}\left(\frac{\Delta}{k^{}_F}\right)^2\left\{
1+\frac{1}{2}\left(\frac{\Delta}{k^{}_F}\right)^2\right\}\,,
\end{equation}
which is due to the distribution of electrons over the whole thin layer ${\cal D}_F$. Since the eigenvalue is the
same for any CPT-like state $|NC,...\hat{\gamma}^{\dag}_{q^{}_{\bf k},\bf k}...\rangle$, then the energy level
$E_{NC}$ has a huge degree of degeneracy, that is equal to $3^{\overline{\delta {\cal N}}/2}$.

As to the construction
$\widehat{\Psi}_{NC}(...\hat{\gamma}^{\dag}_{q^{}_{\bf k},\bf
k}...)$, the occupation of all the thin layer ${\cal D}_F$ in
(\ref{Psi_j}) is dictated by the conservation of particle number.
Indeed, as it follows from (\ref{base}) and (\ref{base1}), the
operators $\hat{\gamma}^{\dag}_{q^{}_{\bf k},\bf k}$ describe the
distribution of two electrons among the four states
$|$$\uparrow$,${\bf k}\rangle$, $|$$\downarrow$,${\bf k}\rangle$,
$|$$\uparrow$,$-{\bf k}\rangle$, $|$$\downarrow$,$-{\bf
k}\rangle$. Because of this, in order to distribute all electrons,
which at the dense packing (into Fermi sphere) were located in the
layer ($k^{}_F$$-\Delta$)$\leq$$k$$\leq$$k^{}_F$, we need in a
doubled volume in the wavevector space. In the case
$\Delta$$\ll$$k^{}_F$ practically the whole thin layer ${\cal
D}_F$ (see in Fig.1a) corresponds to a such double volume, for
which ($k^{}_F$$-\Delta$)$\leq$$k$$\leq$($k^{}_F$$+\Delta$). In
the general case of the construction $\widehat{\Psi}_{NC}$ we can
use arbitrary number of different operators
$\hat{\gamma}^{\dag}_{q^{}_{\bf k},\bf k}$, what can be written in
the form:
\begin{eqnarray}\label{Psi_jl}
&&\widehat{\Psi}_{NC}(...(\hat{\gamma}^{\dag}_{q^{}_{\bf k},\bf k})^{l_{\bf k}}...)=\prod_{{\bf k}\in {\cal
D}_F^{(+)}}(\hat{\gamma}^{\dag}_{q^{}_{\bf k},\bf k})^{l^{}_{\bf k}}\,,\\ &&(q^{}_{\bf k}=0,\pm 1;\;\;l^{}_{\bf
k}=0,1)\,,\nonumber
\end{eqnarray}
where the zero power of an operator equals the unity operator,
i.e. $(\hat{\gamma}^{\dag}_{q^{}_{\bf k},\bf
k})^0$$\equiv$$\hat{{\bf 1}}$ for any ${\bf k}$ and $q^{}_{\bf
k}$.

It should be noted that due to the full spherical symmetry on the translational degrees of freedom (i.e. with
respect to the directions of wavevectors ${\bf k}$) in the operator construction
$\widehat{\Psi}_{NC}(...\hat{\gamma}^{\dag}_{q^{}_{\bf k},\bf k}...)$ (see (\ref{Psi_j})) the CPT-like states
(\ref{NC_j}) correspond to the zero total orbital momentum $L_{total}$=0. While for the states formed with the use
of the more general construction (\ref{Psi_jl}), there exist states with $L_{total}$$\ne$0.

Note also that the ground state in the BCS theory \cite{Bar} can be written in the form
\begin{equation}\label{BCS}
|BCS\rangle=\prod_{{\bf k}}\left\{(1-\eta^{}_{\bf
k})^{1/2}+\eta^{1/2}_{\bf k}\hat{a}^{\dag}_{\uparrow{\bf
k}}\hat{a}^{\dag}_{\downarrow -{\bf k}}\right\}|0\rangle\,,
\end{equation}
where $\eta^{}_{\bf k}$ are variational coefficients. Let us
discuss some properties of the states
$|NC,...\hat{\gamma}^{\dag}_{q^{}_{\bf k},\bf k}...\rangle$, which
are quite different from those of the state $|BCS\rangle$
in the BCS theory:\\
I. $|NC,...\hat{\gamma}^{\dag}_{q^{}_{\bf k},\bf k}...\rangle$
are eigenstates for the particle number operator
$\widehat{N}$=$\sum_{s,{\bf k}}\hat{a}^{\dag}_{s{\bf
k}}\hat{a}^{}_{s{\bf k}}$:
\begin{equation}\label{N}
\widehat{N}|NC,...\hat{\gamma}^{\dag}_{q^{}_{\bf k},\bf
k}...\rangle={\cal N}|NC,...\hat{\gamma}^{\dag}_{q^{}_{\bf k},\bf
k}...\rangle\,.
\end{equation}
II. The states $|NC,...\hat{\gamma}^{\dag}_{q^{}_{\bf k},\bf
k}...\rangle$ are eigenstates for the total momentum operator
$\widehat{\bf P}$=$\sum_{{\bf k},s}(\hbar{\bf
k})\hat{a}^{\dag}_{s{\bf k}}\hat{a}^{}_{s{\bf k}}$. For example,
if the Fermi sphere is constructed around the wavevector ${\bf
K}$, then we have:
\begin{equation}\label{P_NC}
\widehat{\bf P}|NC({\bf K})\rangle={\cal N}(\hbar {\bf
K})|NC({\bf K})\rangle\,,
\end{equation}
where $|NC({\bf K})\rangle$ denotes arbitrary state
$|NC,...\hat{\gamma}^{\dag}_{q^{}_{\bf k},\bf k}...\rangle$. The
consideration above dealt with the particular case ${\bf K}$=0,
but the generalization to arbitrary ${\bf K}$ is almost
elementary and it is achieved by the formal substitution
$\hat{a}^{\dag}_{s,{\bf k}}$$\to$$\,\hat{a}^{\dag}_{s,{{\bf
k}+{\bf K}}}$. The eigenenergy of the states $|NC({\bf
K})\rangle$ is
\begin{equation}\label{E_K}
E_{NC}({\bf K})=E_{NC}+\frac{(\hbar {\bf K})^2}{2m}\,{\cal N}
\end{equation}
i.e. the quadratic in ${\bf K}$ dispersion law takes place.\\
III. The states $|NC,...\hat{\gamma}^{\dag}_{q^{}_{\bf k},\bf
k}...\rangle$ do not depend on the value and sign of the coupling
constant $g$, i.e. they exist in both cases of weak and strong
coupling, and for the case of interparticle repulsion. Although it
should be noted that the particle conservation law leads to a
dependence of the eigenenergy $E_{NC}$ on the other parameter
$\Delta$ (see (\ref{E_NC})).\\
IV. All the states $|NC,...\hat{\gamma}^{\dag}_{q^{}_{\bf k},\bf
k}...\rangle$ belong to the same energy level $E_{NC}$, which,
consequently, has a huge degree of degeneracy, while the ground
state $|BCS\rangle$ in the BCS theory are non-degenerate.

Note, that a possibility of the formation of the triplet state of Cooper pairs has been studied in the papers
\cite{Anderson,Gor'kov,Vaks,Privorotskii,Larkin} for the case of Hamiltonians different from the standard BCS
Hamilotian (\ref{W}).

\section{Magnetism of the delocalized electrons on the base of CPT-like
states}

Evidently, the CPT-like states constitute a special class of
eigenstates of the total Hamiltonian $\widehat{H}_{BCS}$ in view
of the independence on the coupling constant $g$, while,
undoubtedly, there exist other eigenstates with a nontrivial
analytical $g$-dependence of the energy $E$($g$). Because of this
a question about the physical realization of the states
$|NC\rangle$ requires a separate consideration. In the case of
interparticle attraction ($g$$<$0) the energy $E_{NC}$ for the
CPT-like state lies above the ground-state energy of the BCS
theory. However, in the theory with interparticle repulsion
($g$$>$0) it is possible, in principle, that the states
$|NC,...\hat{\gamma}^{\dag}_{q^{}_{\bf k},\bf k}...\rangle$ will
be the ground state, because other states acquire a positive
increment to the energy. Since the states \{$|NC\rangle$\} carry
the macroscopic magnetic moment, then in this case the set of
states \{$|NC\rangle$\} can serve, for example, for the
description of ferromagnetism related to the conductivity
electrons in metals.

In particular, under some simplifying assumptions on the
dispersion law and interaction Hamiltonian (see in Appendix) it
can be proved rigorously, that, indeed, the states
\{$|NC\rangle$\} have the lowest energy at $g$$>$0. Basing on the
subspace of the states \{$|NC\rangle$\}, we can now construct a
model describing the paramgnetism and ferromagnetism in an
ensemble of fermions. First of all, in the framework of our
approach we describe the induced magnetization (i.e. the
paramagnetism) under the action of external magnetic field. To do
this we will consider a simplified thermodynamic model of the
particle ensemble, which is described by the states
$|NC,...\hat{\gamma}^{\dag}_{q^{}_{\bf k},\bf k}...\rangle$ only,
in a static magnetic field ${\bf B}$=$B$${\bf e}_z$. The
quantization axis $z$ is directed along ${\bf B}$. This model
corresponds to the ensemble consisting of $\overline{\delta {\cal
N}}/2$ particles with the spin $S$=1, when every particle can
carry the magnetic moment $2q\mu^{}_B$ depending on the spin
projection $\hbar q$ with respect to the $z$ axis ($q$=0,$\pm 1$).
In the linear approximation on the magnetic field the states of
vector particles acquire the energy shifts $2\mu^{}_B q B$.

As is known, in the equilibrium thermodynamic ensemble of non-interacting particles with the spin $j$ in the
magnetic field ${\bf B}$ the magnetic moment ${\bf M}$ along the vector ${\bf B}$ is formed. Its value in the
thermodynamic limit ($V$$\to$$\infty$) is calculated as:
\begin{equation}\label{M}
M=\frac{\sum_{m^{}_j=-j}^{+j}(m^{}_j/j)\exp\left\{\frac{2\mu^{}_Bm^{}_jB}{k^{}_BT}\right\}}
{\sum_{m^{}_j=-j}^{+j}\exp\left\{\frac{2\mu^{}_Bm^{}_jB}{k^{}_BT}\right\}}\,M_0\,,
\end{equation}
where $M_0$ is the maximal magnetic moment. In the case under
consideration $j$=1, that leads to
\begin{equation}\label{M1}
M=\frac{\exp\left\{\frac{2\mu^{}_BB}{k^{}_BT}\right\}-\exp\left\{-\frac{2\mu^{}_BB}{k^{}_BT}\right\}}
{1+\exp\left\{\frac{2\mu^{}_BB}{k^{}_BT}\right\}+\exp\left\{-\frac{2\mu^{}_BB}{k^{}_BT}\right\}}\,M_0
\end{equation}
with the maximal magnetic moment:
\begin{equation}\label{M0}
M_0=2\mu^{}_B\,\overline{\delta {\cal N}}/2=\mu^{}_B\,\overline{\delta {\cal N}}\,,
\end{equation}
which is achieved at $T$=0. In this case the spins of all vector particles are directed along the magnetic field,
i.e. the ensemble is in the state $|NC,...\hat{\gamma}^{\dag}_{-1,\bf k}...\rangle$ formed by the operator
construction:
\begin{equation}\label{Psi_1}
\widehat{\Psi}_{NC}(...\hat{\gamma}^{\dag}_{-1,\bf k}...)=\prod_{{\bf k}\in {\cal
D}_F^{(+)}}\hat{\gamma}^{\dag}_{-1,\bf k}\,.
\end{equation}
Note, that the magnetic moment of arbitrary state
$|NC,...\hat{\gamma}^{\dag}_{q^{}_{\bf k},\bf k}...\rangle$ is
equal to
\begin{equation}\label{M_j}
M(...\hat{\gamma}^{\dag}_{q^{}_{\bf k},\bf k}...)=2\mu^{}_B\sum_{{\bf k}\in {\cal D}_F^{(+)}}q^{}_{\bf k}
\end{equation}
and it can takes the values $-M_0$,($-M_0$$+2\mu_B$),...,
($M_0$$-2\mu_B$),$\,M_0$.

\begin{figure}[t]
\centerline{\scalebox{0.5}{\includegraphics{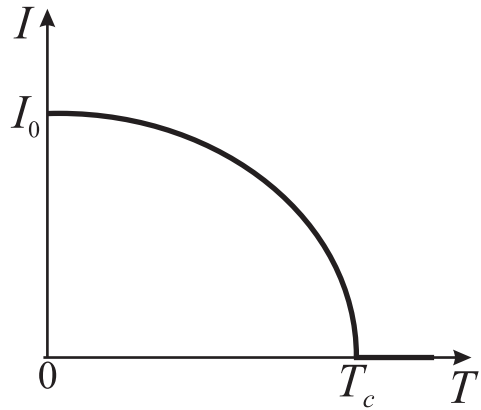}}}\caption{
Typical temperature dependence of the spontaneous magnetization
$I$($T$) defined as a solution of the equation (\ref{I}).}
\end{figure}

Now we describe in the framework of our approach (i.e. remaining
in the basis of states \{$|NC\rangle$\}) a model of the
ferromagnetism of electron gas. For this purpose, we introduce the
additional Hamiltonian of the spin-spin interaction
$\widehat{H}_{ss}$. The presence of this additional Hamiltonian
leads (at certain conditions) to the spontaneous magnetization,
when the states with non-zero macroscopic magnetic moment become
energetically preferable.

Then, for the sake of simplicity we will use the standard method of molecular (or mean) field, assuming the
existence of the internal magnetic field ${\bf B}_{mol}$=$\lambda$${\bf I}$, which is proportional to the
magnetization ${\bf I}$=${\bf M}$$/V$, and $\lambda$ is the Weiss coefficient. In the framework of this approach
the Hamiltonian $\widehat{H}_{ss}$ can be written as:
\begin{equation}\label{H_ss0}
\widehat{H}_{ss}=-2\mu^{}_B\lambda({\bf I}\cdot\widehat{\bf S}^{(s)})\,,
\end{equation}
where $\widehat{\bf S}^{(s)}$ is the spin operator of the ensemble of fermions.
Choosing the quantization axis  $z$ along the magnetization vector ${\bf I}$,
the expresion for  (\ref{H_ss0}) can be rewritten in the following way:
\begin{equation}\label{H_ss}
\widehat{H}_{ss}=-\frac{4G_{ss}}{V\hbar^2}\langle\widehat{S}^{(s)}_z\rangle\widehat{S}^{(s)}_z\,,
\end{equation}
where $G_{ss}$=$\hbar^2$$\mu^2_B$$\lambda$ is the interaction constant, and $\widehat{S}^{(s)}_z$ is the
$z$-projection of the spin operator.

The presence of the mean field  ${\bf B}_{mol}$ in combination with (\ref{M1}) leads to the equation
\begin{equation}\label{I}
I=\frac{\exp\left\{\frac{2\mu^{}_B{\lambda}I}{k^{}_BT}\right\}-\exp\left\{-\frac{2\mu^{}_B{\lambda}I}{k^{}_BT}\right\}}
{1+\exp\left\{\frac{2\mu^{}_B{\lambda}I}{k^{}_BT}\right\}+
\exp\left\{-\frac{2\mu^{}_B{\lambda}I}{k^{}_BT}\right\}}\,I_0\,.
\end{equation}
The spontaneous magnetization $I$=$|{\bf I}|$ as a function of the temperature $T$ is found from the solution of
this equation. The typical dependence $I$($T$) is shown in Fig.2. The maximal magnetization $I_0$ is equal to:
\begin{equation}\label{I_0}
I_0=\mu^{}_B n\,\frac{\overline{\delta {\cal N}}}{{\cal N}}\,.
\end{equation}
It should be stressed that though the result (\ref{I}) coincides formally with the mean field result for the
Heisenberg ferromagnetic with localized magnetic ions, but in our case we deal with the magnetic ordering of the
collective conductivity electrons (if one keep in mind the description of the magnetism in metals). Note also,
that the magnetic ordering of the conductivity electrons, according to our model, automatically leads to the
small fractional value

\begin{equation}\label{I01}
\frac{I_0}{\mu^{}_B n}=\frac{\overline{\delta {\cal N}}}{{\cal N}}\ll 1\,,
\end{equation}
i.e. of the magnetic moment in the $\mu^{}_B$ units per one electron.

As it follows from (\ref{I}), at $T$=0 the spontaneous
magnetization does exist (and it equals to $I_0$) for arbitrary
value of the coefficient $\lambda>0$ (formally it can be as small
as one likes), and, consequently, for arbitrary value of the
interaction constant $G_{ss}$$>$0 in (\ref{H_ss}). This
circumstance is connected with the specific dispersion law
(\ref{H_01})-(\ref{dN1}), which leads automatically to the fact
that the energy of CPT-like states is the lowest energy (see
Appendix). This fact, in its turn, allows us to construct the
above model of the magnetism in the ensemble of fermions, based on
the basis of states  \{$|NC\rangle$\}.

However for the more realistic quadratic dispersion law (\ref{k2})
we can not use the formulas given above for arbitrary
$\lambda$$>$0 ($G_{ss}$$>$0), because the energy $E_{NC}$ lies
above the energy of the Fermi-sphere $E_{F}$ (see
(\ref{E_NC})-(\ref{DE_NC})), i.e. the states  \{$|NC\rangle$\} is
not the lowest energy states (for Hamiltonian
$\widehat{H}_{BCS}$). Therefore to justify our approach it is
necessary that the negative contribution to the energy of some
CPT-like states due to the spin-spin interaction (\ref{H_ss})
compensates the positive additional term in the kinetic energy
$\Delta E_{NC}$ (see (\ref{DE_NC})).  In this case the energy of
such CPT-like states will be lower than the energy of the
Fermi-sphere $E_F$, which gives us the ground to use the basis
\{$|NC\rangle$\} in the description of the magnetism for the
quadratic dispersion law too.

It is obvious that the state $|NC,...\hat{\gamma}^{\dag}_{-1,\bf k}...\rangle$, formed by the construction
(\ref{Psi_1}), where the spins of all the vector Cooper pairs are oriented in the same direction, acquires the
maximal value of the negative addition to the energy due to the spin-spin interaction. This state has the largest
spin moment $S_z$=$\hbar\overline{\delta {\cal N}}$/2 and the corresponding negative correction to the energy
$H^{(m)}_{ss}$$<$0 (at zero temperature $T$=0) is equal to:
\begin{eqnarray}\label{DE_ss}
&& H^{(m)}_{ss}\equiv\langle NC,...\hat{\gamma}^{\dag}_{-1,\bf
k}...|\widehat{H}_{ss}|NC,...\hat{\gamma}^{\dag}_{-1,\bf k}...\rangle
=\nonumber \\
&&-\frac{G_{ss}\overline{\delta {\cal N}}^{\,2}}{V}=-9\,G_{ss}\left(\frac{\Delta}{k^{}_F}\right)^2 n{\cal N}\,,
\end{eqnarray}
where the particle density $n$ can be expressed in terms of the
radius of the Fermi-sphere as $n$=$k_F^3$/$3\pi^2$ (in the case of
the quadratic dispersion law). In order to the absolute value of
the negative correction (\ref{DE_ss}) exceeds the energy addition
(\ref{DE_NC}), the following condition should be satisfied:
\begin{equation}\label{g_rho}
G_{ss}\rho^{}_F > 1\,,
\end{equation}
where the parameter $\rho^{}_F$=$m$$k^{}_F$/$\pi^2$$\hbar^2$ is the state density near the Fermi surface.
In this case the inequality takes place
\begin{equation}\label{<}
E_{NC}+H^{(m)}_{ss}<E_F
\end{equation}
and, due to this reason, (\ref{g_rho}) can be considered as a criterion of the applicability of our model, when
for the description of magnetic ordering near $T$=0 we can use the set of wavefunctions \{$|NC\rangle$\}.

It should be noted that (\ref{g_rho}) formally coincides with the criterion of ferromagnetism in the Stoner model
\cite{Stoner}. This circumstance is connected with the use of the state $|NC,...\hat{\gamma}^{\dag}_{-1,\bf
k}...\rangle$ in the deduction of (\ref{g_rho}). From the other hand, namely this state describe the case, when
the spins of all fermions in the spherical layer  $k_F$$-$$\Delta$$\le$$k$$\le$$k_F$$+$$\Delta$ are oriented
along the same direction, that, in its turn, corresponds to two different (i.e. with different radiuses)
Fermi-spheres. Indeed, for particles with the spin up ($\uparrow$) the corresponding radius of Fermi-sphere is
$k_{F\uparrow}$=$k_F$$-$$\Delta$, and for particles with the spin down ($\downarrow$) we have
$k_{F\downarrow}$=$k_F$$+$$\Delta$ (i.e. $k_{F\downarrow}$$\ne$$k_{F\uparrow}$). But namely similar approach is
used in the description of the ferromagnetism in the Stoner model.

Despite the formal identity of the criterion (\ref{g_rho}) to the
Stoner criterion of ferromagnetism \cite{Stoner}, our approach has
several principal distinctions. For example, in our approach the
magnetism is governed first of all by the redistribution of
fermions within the thin layer ${\bf k}$$\in$${\cal D}_F$ near the
Fermi-sphere. This leads automatically to the small value of the
magnetic moment (\ref{I01}) per one particle (for instance, per
one conductivity electron in metal). This result as whole does not
contradict to the experimental data, according to which for the
overwhelming majority of metallic magnetics the magnetic moment,
originating from the conductivity electrons, does not exceed few
percent of the value  $\mu^{}_B$ per one free electron. Note also
that the presence of the thin layer ${\cal D}_F$ in the ensemble
of fermions is caused, according to our approach (i.e. in the
framework of the general BCS ideology), by the interaction
(scattering) with some other particles (for example, photons,
phonons, excitons etc.). Apart from this, the existence of such
thin layer can be connected with the presence of the energy gap
near the Fermi-surface.

However the main distinguishing feature of our approach is
connected with the description of the magnetic properties of
conductivity electrons in the framework of notion of a gas of
particles with the spin 1. In this case we can use the CPT-like
states with the non-zero total momentum  $|NC({\bf K})\rangle$
also, i.e. when the Fermi-sphere is constructed around the
non-zero wavevector ${\bf K}$ (the momentum and energy of these
states are defined by (\ref{P_NC}) and (\ref{E_K})). Since the
states $|NC({\bf K})\rangle$ carry the magnetic moment, then these
states can be interpreted as spin waves for the conductivity
electrons. Moreover, since the CPT-like states are constructed
according to the type of free vector particles, then a notion of a
gas of Bose-particles, which are Cooper pairs with spin 1 and
electric charge $2e$, emerges naturally in the description of
magnetic properties. This conception can be realized by the
introduction of the operators of creation of particles
$\hat{\gamma}^{\dag}_{q}({\bf K})$ with the spin projection
$\hbar$$q$ ($q$=0,$\pm 1$) and wavevector ${\bf K}$. Due to the
quadratic in  ${\bf K}$ dispersion law in (\ref{E_K}), the
dispersion law for the vector particles is naturally presented as
quadratic:
\begin{equation}\label{e}
\tilde{\varepsilon}({\bf K})=\frac{(\hbar{\bf K})^2}{2\widetilde{m}}\,,
\end{equation}
where $\widetilde{m}$ is the effective mass of the pair. These
Bose-particles can form a base of conception of the magnons (i.e.
on the elementary excitations of spin waves) with the spin $S$=1
in the subsystem of conductivity electrons. Also there is, in
principle, the possibility of the Bose-Einstein condensation.

Our model can be extended to the case, allowing an ordering of the electron angular moments (spins) of the
moveless ions in solids and their spin interaction with the conductivity electrons. We will describe this
interaction by the Hamiltonian $\widehat{H}_{fs}$:
\begin{equation}\label{H_fs}
\widehat{H}_{fs}=\frac{A_{fs}}{V\hbar^2}\,(\widehat{\bf S}^{(f)}\cdot\widehat{\bf S}^{(s)})\,,
\end{equation}
where $\widehat{\bf S}^{(f)}$ is the spin operator of the localized electrons (i.e. of the moveless ions), and
$A_{fs}$ is the interaction constant. The total spin of one ion is denoted as $J_f$, and the spatial density of
ions is equal to $n_f$.

In the model of mean field the Hamiltonian (\ref{H_fs}) can be rewritten in the form
\begin{equation}\label{H_fs1}
\widehat{H}_{fs}=\frac{A_{fs}}{V\hbar^2}\,(\langle\widehat{S}^{(f)}_z\rangle\cdot\widehat{S}^{(s)}_z)\,,
\end{equation}
where the axis $z$ is directed along the vector of mean spin of ions ${\bf S}^{(f)}$=$\langle\widehat{\bf
S}^{(f)}\rangle$.

The maximal negative energy term  $H^{(m)}_{fs}$$<$0 due to the Hamiltonian (\ref{H_fs1}) for the set of
CPT-states \{$|NC\rangle$\} is achieved in the case when the spins of all ions are oriented in the same directions
($\langle\widehat{S}^{(f)}_z\rangle$=$\hbar$$J_f$$n_fV$), and the spins of all vector pairs are oriented parallel
(at  $A_{fs}$$<$0) to the spin of ions, or antiparallel (at $A_{fs}$$>$0) to the spin of ions. Take for the
specificity $A_{fs}$$>$0. In this case
\begin{eqnarray}\label{Hmfs}
H^{(m)}_{fs}=&&\langle NC,...\hat{\gamma}^{\dag}_{-1,\bf k}...|\widehat{H}_{fs}|NC,...\hat{\gamma}^{\dag}_{-1,\bf
k}...\rangle= \nonumber\\
&& -A_{fs}J_f n_f \overline{\delta {\cal N}}/2\,.
\end{eqnarray}
Now the more general (with respect to (\ref{<})) condition of applicability of our model at $T$=0 can be written as
\begin{equation}\label{<fs}
E_{NC}+H^{(m)}_{ss}+H^{(m)}_{fs}<E_F\,,
\end{equation}
which leads to
\begin{equation}\label{g_rho_fs}
\left(\frac{J_f}{6}
\frac{n_f}{n}\frac{k_F}{\Delta}\,|A_{fs}|+G_{ss}\right)\rho^{}_F >
1\,.
\end{equation}
This inequality due to the condition $k_F/\Delta\gg 1$ can be
satisfied for sufficiently small values  $A_{fs}$ even at
$G_{ss}$=0.

As a whole, the described above approach to the magnetism in the ensemble of fermions corresponds to the following
formal scheme. Let us consider the Hamiltonian of general form:
\begin{equation}\label{H_gen}
\widehat{H}=\widehat{H}_{BCS}+\widehat{\cal H}_{(S)}\,,
\end{equation}
where the Hamiltonian $\widehat{\cal H}_{(S)}$ contains the interactions with the spin of particles, i.e. the
interactions connected with the magnetism. For example:
\begin{equation}\label{H_S}
\widehat{\cal H}_{(S)}=\widehat{H}_{ss}+\widehat{H}_{fs}+2\mu^{}_B({\bf B}\cdot\widehat{\bf S}^{(s)})\,.
\end{equation}
The Hamiltonian $\widehat{H}_{BCS}$ in (\ref{H_gen}) plays a role of the basic Hamiltonian, and the operator
$\widehat{\cal H}_{(S)}$ is considered as a perturbation. Then, in the linear approximation we find the maximal
negative correction $H_{(S)}^{(m)}$$<$0 to the energy of CPT-like states:
\begin{equation}\label{H_Sm}
H_{(S)}^{(m)}=\langle NC,...\hat{\gamma}^{\dag}_{-1,\bf k}...|\widehat{\cal
H}_{(S)}|NC,...\hat{\gamma}^{\dag}_{-1,\bf k}...\rangle\,.
\end{equation}
From consideration of the minimality of energy we find the criterion of applicability of our model at $T$=0:
\begin{equation}\label{Krit}
E_{NC}+H_{(S)}^{(m)} < E_F\,.
\end{equation}
The inequalities (\ref{<}) and (\ref{<fs}) (correspondingly, (\ref{g_rho}) and (\ref{g_rho_fs})) should be
considered as some particular cases of the general inequality (\ref{Krit}). Note, the spin Hamiltonian
$\widehat{\cal H}_{(S)}$ in (\ref{H_gen}) does not influence (in the linear approximation) on the energy $E_F$ of
the Fermi state $|F\rangle$ (see (\ref{Fermi})), because this state correspond to the zero total spin (and
orbital angular momentum).

It should be noted, that in the framework of described approach
one can consider (at least formally) also the case of
interparticle attraction, i.e. $g$$<$0 in (\ref{W}). However in
this case the ground state energy of the BCS theory $E_{BCS}$ (see
\cite{Bar}) lies below the Fermi energy, i.e. $E_{BCS}$$<$$E_{F}$.
Due to this reason at $g$$<$0 the criterion of applicability of
our model of the magnetism, based on the CPT-like states, differs
from (\ref{Krit}) and has the following form:
\begin{equation}\label{Krit_2}
E_{NC}+H_{(S)}^{(m)} < E_{BCS}\,.
\end{equation}

\section{CPT-like state and superfluidity in $^3$He}

Apart from the delocalized electrons in metals another object, to which the obtained results can be related, is
liquid $^3$He. Atoms of $^3$He are fermions with the spin 1/2 and due to this reason the use of the standard BSC
Hamiltonian (\ref{W}) for theoretical description is well-grounded (at lest, on the qualitative level).

As is known (see, for example, the reviews \cite{Mineev,Volovik}
and references therein), the superfluid phase of $^3$He is
characterized by the formation of Cooper pairs with the spin
$S$=1. This fact, according to our approach, can be interpreted as
a consequence of the repulsion ($g$$>$0) caused by the $s$-wave
scattering, when the vector pairing can be energetically
favorable. Note that the $s$-wave scattering is included in the
Hamiltonian (\ref{W}).

This fact, according to our approach, can be interpreted, for
example, as a consequence of the effective repulsion between
quasiparticles in liquid $^3$He, i.e. when $g$$>$0 and the vector
pairing can be energetically favorable.

Note, that the operator constructions of general form (\ref{Psi_jl}) allow the formation of spherically
asymmetrical states $|NC\rangle$, i.e. the states with the non-zero orbital angular momentum. Therefore there
exists a possibility to describe paired states, which are characterized not only by the spin $S$=1, but also by
the orbital momentum  $L$$\neq$0 and, in particular, $L$$=$1. Here we deal with the angular orbital momentum of
the relative motion of particles in the pair. The orbital momentum connected with the translational motion of the
Cooper pairs is described by the states  $|NC({\bf K})\rangle$ with the non-zero total linear momentum (see
(\ref{P_NC}) иand(\ref{E_K})) and by their coherent superpositions.

Thus, our approach can be considerably easily inserted into the existed general picture of the theoretical
description of $^3$He based on the notion of Cooper pairs with $S$=1 and $L$$=$1. Basing on this notion, we can
now introduce additional interaction Hamiltonians (between vector pairs, spin-orbit coupling, spin-spin, with
magnetic field etc.), and allow for the more detail description of physical properties (for instance, the
classification of superfluid phases in $^3$He-$B$, $^3$He-$A$, and $^3$He-$A_1$). In other words, the obtained
CPT-like states can be considered, in a formally consistent way, as the zero approximation (corresponding to the
spin $S$=1 of Cooper pairs) to the standard theoretical scheme.

Let us add that since the BCS model is used in the description of neutron stars (see in the review \cite{Ginz}),
then the substance of neutron stars can be, in principle, an object of application of the obtained results
(including the magnetism).

\section{Conclusion}

In the framework of standard mathematical BCS model we have
considered the ensemble of interacting fermions with the spin 1/2.
Usually this model is used for the description of Cooper pairs
with the total spin $S$=$\,0$. However, as it turns out, the
standard BCS Hamiltonian describes also the vector ($S$=1) pairing
of particles with the opposite linear momenta close to the Fermi
sphere. Thus, it is shown that, in principle, for the description
of vector Cooper pairs with the spin 1 it is not necessary to
introduce into the interaction Hamiltonian the corresponding
vector operator constructions, i.e. it is possible to remain in
the framework of the operator (\ref{W}), formed only by the scalar
constructions (\ref{aa}). The found states nullify the interaction
Hamiltonian (\ref{W}) and, therefore,  they have some analogy with
the known CPT effect. Moreover, there are good reasons to believe
that at certain conditions these CPT-like states can belong to the
lower part of energy spectrum, and, consequently, play a
significant role in the description of physical properties of the
given ensemble.

Since the CPT-like states have a huge degree of degeneracy and
carry the macroscopic magnetic moment, then it is logical to apply
them to the description of the magnetic ordering for the
delocalized electrons in the case $g$$>$0, when the vector pairing
is energetically favorable. In particular, we have proposed a new
approach to the explanation of the ferromagnetism (and the
magnetic ordering in general) connected with the delocalized
electrons (i.e. itinerant magnetism). Moreover, here the concept
of magnons with the spin $S$=1 in the subsystem of the delocalized
electrons naturally emerges. Apart from the magnetism in metals
(conductors), the obtained results may have a potential
significance for the description of the superfluidity in $^3$He,
explaining, for example, the vector type ($S$=1) of Cooper pairs
as a consequence of the repulsion ($g$$>$0) in the interaction
Hamiltonian (\ref{W}).

Thus, there are serious reasons to assume that the standard mathematical BCS model is more universal and, apart
from the superconductivity (at $g$$<$0) it can serve as a base in the description of other affects in metals
(conductors) and in  quantum Fermi liquids (for example, at $g$$>$0).

The presented results on the magnetism in metals and the
superfluidity in $^3$He are preliminary (discussional) and have a
character of a review of some possible consequences under the
assumption that the found CPT-like states have a concrete physical
sense and they can belong to the lower part of energy spectrum.
However, even such a qualitative analysis confirms that the
proposed approach deserves an attention and further development.

Authors thank G.I. Surdutovich, E.G. Batyev, L.V. Il'ichev, and
A.M. Tumaikin for useful discussions. This work was supported by
grants INTAS-SBRAS (06-1000013-9427), RFBR (07-02-01230,
07-02-01028, 08-02-01108), and Presidium SB RAS.

\appendix
\section{}

\renewcommand{\theequation}{\thesection\arabic{equation}}

Let us show that at some simplifying assumptions on the operators
(\ref{H_0}) and (\ref{W}) the energy level $E_{NC}$ is the ground
level at $g$$>$0.

So in the interaction Hamiltonian (\ref{W}) we will assume the equality  $G$$({\bf k}_1,$${\bf k}_2)$=1 for the
formfactor. Then the operator $\widehat{W}$ can be written in the form:
\begin{eqnarray}\label{W1}
&&\widehat{W}=g\,\widehat{B}^{\dag}\widehat{B}\,,\\
&&\widehat{B}=\frac{1}{\sqrt{V}}\sum_{{\bf k}\in {\cal D}_F} \hat{a}^{}_{\downarrow -{\bf k}}\hat{a}^{}_{\uparrow
{\bf k}}\,.\nonumber
\end{eqnarray}
Another approximation is related to the dispersion law
$\varepsilon^{}_{\bf k}$ in the unperturbed Hamiltonian
(\ref{H_0}). for the thin spherical layer ${\cal D}_F$ (when
$\Delta$$\ll$$k^{}_F$) the energy of particles
$\varepsilon^{}_{\bf k}$ with different ${\bf k}$ are almost the
same and, due to this reason, we will assume their equality
$\varepsilon^{}_{\bf k}$=$\bar{\varepsilon}$ for all ${\bf
k}$$\in$${\cal D}_F$. In this case the kinetic energy operator
$\widehat{H}_0$ is split into the two summand:
\begin{equation}\label{H_01}
\widehat{H}_0=\bar{\varepsilon}\,\widehat{\delta {\cal N}}+\widehat{H}^{(\Phi)}_0\,,\\
\end{equation}
Here the first summand corresponds to the states with
$(k^{}_F$$-$$\Delta)$$<$$|{\bf k}|$$<$$(k^{}_F$$+$$\Delta)$ and
it is proportional to the operator of the particle number in the
spherical layer ${\cal D}_F$:
\begin{equation}\label{dN1}
\widehat{\delta {\cal N}}=\sum_{s,{\bf k}\in{\cal D}_F}\hat{a}^{\dag}_{s{\bf k}}\hat{a}^{}_{s{\bf k}}\,.
\end{equation}
The dispersion law $\varepsilon^{}_{\bf k}$ for $|{\bf
k}|$$<$$(k^{}_F$$-$$\Delta)$ in the second summand
$\widehat{H}^{(\Phi)}_0$ in (\ref{H_01}) is assumed arbitrary:
\begin{equation}\label{H_Phi}
\widehat{H}^{(\Phi)}_0=\sum_{s,|{\bf k}|<(k^{}_F-\Delta)}\varepsilon^{}_{\bf k}\,\hat{a}^{\dag}_{s{\bf
k}}\hat{a}^{}_{s{\bf k}}\,.
\end{equation}
Take arbitrary eigenvector $|\Psi\rangle$ for the total Hamiltonian
$\widehat{H}_{BCS}$=$\widehat{H}_0$$+$$\widehat{W}$, i.e. $\widehat{H}_{BCS}$$|\Psi\rangle$=$E$$|\Psi\rangle$.
Then, in view of (\ref{H_01}) and (\ref{W1}), for the eigenenergy $E$ the following relationships is fulfilled:
\begin{eqnarray}\label{E}
&&E=\langle\Psi |\widehat{H}_{BCS}|\Psi\rangle=\langle\Psi |\widehat{H}_0|\Psi\rangle+g\,\langle\Psi |\widehat{B}^{\dag}\widehat{B}|\Psi\rangle = \nonumber\\
&& \bar{\varepsilon}\,\langle\Psi |\widehat{\delta {\cal N}}|\Psi\rangle +\langle\Psi
|\widehat{H}^{(\Phi)}_0|\Psi\rangle+g\,\langle\Psi |\widehat{B}^{\dag}\widehat{B}|\Psi\rangle\,.
\end{eqnarray}
By the problem statement we are interested only in such states,
which can be symbolically presented as:
\begin{equation}\label{Psi_G}
|\Psi\rangle=\widehat{\Psi}\,\widehat{\Phi}(\Delta)\,|0\rangle\,,\quad \langle\Psi |\Psi\rangle=1\,,
\end{equation}
where some operator $\widehat{\Psi}$ acts only on the states in
the spherical layer ${\bf k}$$\in$${\cal D}_F$, and the fixed
construction $\widehat{\Phi}$($\Delta$) in accordance with
(\ref{Ferm1}) describes the fully occupied states with $|{\bf
k}|$$<$$(k^{}_F$$-$$\Delta)$. Since the kinetic energy operator
(\ref{H_Phi}) also acts only on the states with $|{\bf
k}|$$<$$(k^{}_F$$-$$\Delta)$, then for arbitrary vector
$|\Psi\rangle$ the second summand in (\ref{E}) is fixed:
\begin{equation}\label{E_Phi}
\langle\Psi |\widehat{H}^{(\Phi)}_0|\Psi\rangle=E_{\Phi}\,.
\end{equation}
Apart from this, only the states with conserved number of
particles (at least in average) are physically significant, i.e.
the average of the particle number operator
$\widehat{N}$=$\sum_{s,{\bf k}}\hat{a}^{\dag}_{s{\bf
k}}\hat{a}^{}_{s{\bf k}}$ is equal to the fixed number ${\cal N}$:
\begin{equation}\label{NN}
\langle\Psi |\widehat{N}|\Psi\rangle={\cal N}\,.
\end{equation}
From the other hand, the construction  $\widehat{\Phi}$($\Delta$)
determines the fixed number ${\cal N}_{\Phi}$ of particles,
occupying the sphere with $|{\bf k}|$$<$$(k^{}_F$$-$$\Delta)$.
Consequently, all the physically significant states $|\Psi\rangle$
should also conserve the particle number $\overline{\delta {\cal
N}}$ in the spherical layer ${\cal D}_F$, because ${\cal
N}$=${\cal N}_{\Phi}$$+$$\overline{\delta {\cal N}}$. As a result
we have:
\begin{equation}\label{dN2}
\langle\Psi |\widehat{\delta {\cal N}}|\Psi\rangle=\overline{\delta {\cal N}}\,.
\end{equation}
Using (\ref{E_Phi}) and (\ref{dN2}), the expression  (\ref{E}) can
be rewritten in the following form:
\begin{equation}\label{E1}
E=\bar{\varepsilon}\,\overline{\delta {\cal N}} +E_{\Phi}+g\,\langle\Psi
|\widehat{B}^{\dag}\widehat{B}|\Psi\rangle\,.
\end{equation}
Consider now the case $g$$>$0. From the obvious inequality
\begin{equation}\label{geq}
\langle\Psi |\widehat{B}^{\dag}\widehat{B}|\Psi\rangle=\langle\widehat{B}\Psi|\widehat{B}\Psi\rangle\geq 0\,,
\end{equation}
it follows that
\begin{equation}\label{Egeq}
 E\geq\bar{\varepsilon}\,\overline{\delta {\cal N}} +E_{\Phi}\,.
\end{equation}
From the other hand, since the CPT-like states nullify the
interaction operator $\widehat{W}$, then for the energy $E_{NC}$
we get:
\begin{equation}\label{E_NC2}
E_{NC}=\bar{\varepsilon}\,\overline{\delta {\cal N}} +E_{\Phi}\,.
\end{equation}
Thus, the states $|NC\rangle$ belong to the ground energy level,
as we wished prove.

Moreover, let us show that the energy of all other states lie
above the energy $E_{NC}$. Indeed, consider the eigenstate
$|\Psi\rangle$, for which the condition of nullification of the
operator $\widehat{W}$ is not fulfilled, i.e.
$\widehat{W}$$|\Psi\rangle$$\neq$0. In accordance with (\ref{W1})
this means $\widehat{B}$$|\Psi\rangle$$\neq$0. Then the rigorous
inequality is fulfilled
\begin{equation}\label{geq}
\langle\Psi |\widehat{B}^{\dag}\widehat{B}|\Psi\rangle=\langle\widehat{B}\Psi|\widehat{B}\Psi\rangle > 0\,,
\end{equation}
which, in its turn, leads to the rigorous inequality
$E$$>$$E_{NC}$.

Vice versa, in the case $g$$<$0 it can be proved analogously that the energy $E_{NC}$ is maximal.

It should be noted that although the model dispersion law for free
particles in (\ref{H_01}) is approximate, nevertheless, the
performed analysis argues that our statement that the vector
pairing ($S$=1) is possibly energetically favorable for $g$$>$0 is
not unfounded. Vice versa, in the case of interparticle attraction
($g$$<$0) the vector type of pairing is not energetically
favorable.

Let us add that above we talk only about the states $|NC\rangle$, because we know their exact analytical form and
energy $E_{NC}$. It is well to bear in mind that in the general case there exist a whole family of allied energy
levels, which will be denoted as $\{NC\}$. For example, if in the operator construction $\widehat{\Psi}_{NC}$
(see (\ref{Psi_j})) we replace several operators  $\hat{\gamma}^{\dag}_{q^{}_{\bf k},{\bf k}}$ by some other
constructions, consisting of $\hat{a}^{\dag}_{s,{\bf k}}$ (conserving the total particle number ${\cal N}$), then
as a result we will generate some state $|\Psi\rangle$, which formally does not obey the condition (\ref{W_NC1}),
i.e. $\widehat{W}$$|\Psi\rangle$$\neq$0. At the same time, it is obvious that the state $|\Psi\rangle$ does not
practically differ from the state $|NC\rangle$ (especially in the limit ${\cal N}$$\to$$\infty$). The definition
of the family (sub-band) of levels $\{NC\}$ is a subject of special consideration. Generally speaking the states
from $\{NC\}$ should be characterized by the predominance of Cooper pairs with the spin $S$=1, and by the
considerably weak influence of the operator $\widehat{W}$ on them. It should be noted, that the similar situation
appears in the standard CPT effect in the resonant interaction of light with multilevel atoms. In this case there
are one or several exact CPT-states and a whole family of allied states (with similar physical properties and
close energy). Thus, the exact CPT-states play a role of a kernel-type of special subsystem of levels.

\end{document}